\documentclass[12pt]{article}

\usepackage[style=nature]{biblatex}
\usepackage{graphicx}
\usepackage{bm}
\usepackage{lscape}
\usepackage{amsmath}
\usepackage{braket}
\usepackage{longtable}
\usepackage{array}
\usepackage{xcolor} 
\usepackage{times}
\usepackage{epstopdf}
\usepackage{amssymb}
\usepackage{lineno}

\title{Mid-infrared trace detection\\ with parts-per-quadrillion quantitation accuracy: \\Expanding frontiers of radiocarbon sensing}

\author
{Jun Jiang,$^{1\ast}$ A. Daniel McCartt$^{1\ast}$\\
\\
\normalsize{$^{1}$Center for Accelerator Mass Spectrometry, Lawrence Livermore National Laboratory,}\\
\normalsize{Livermore, California 94550, USA}\\
\normalsize{$^\ast$To whom correspondence should be addressed: jiang12@llnl.gov, mccartt1@llnl.gov.}
}

\date{}

\begin{document} 


\baselineskip24pt

\maketitle 



\begin{abstract}
\baselineskip18pt
Detection sensitivity is one of the most important attributes to consider during selection of spectroscopic techniques. However, high sensitivity alone is insufficient for spectroscopic measurements in spectrally congested regions. Two-color cavity ringdown spectroscopy (2C-CRDS), based on intra-cavity pump-probe detection, simultaneously achieves high detection sensitivity and selectivity. The technique enables mid-infrared detection of radiocarbon dioxide ($^{14}$CO$_2$) molecules in room-temperature CO$_2$ samples, with better than 10 parts-per-quadrillion (ppq, 10$^{15}$) quantitation accuracy (4 ppq on average). These \textit{{highly-reproducible}} measurements, which are the most \textit{{sensitive}} and \textit{{quantitatively accurate}} in the mid-infrared, are accomplished despite the presence of \textit{{orders-of-magnitude}} stronger, one-photon signals from other CO$_2$ isotopologues. This is a major achievement in laser spectroscopy. A room-temperature-operated, compact, and low-cost 2C-CRDS sensor for $^{14}$CO$_2$ benefits a wide range of scientific fields that utilize $^{14}$C for dating and isotope tracing, most notably atmospheric $^{14}$CO$_2$ monitoring to track CO$_2$ emissions from fossil fuels. The 2C-CRDS technique significantly enhances the general utility of high-resolution mid-infrared detection for analytical measurements and fundamental chemical dynamics studies.
\end{abstract}

\vfill     {\centering LLNL-JRNL-850018\par}




Quantifying light absorption is one of the most commonly used strategies to determine the concentration, transition frequencies, and transition cross sections for an analyte of interest. The most sensitive laser absorption techniques invariably utilize an optical cavity~\cite{Romanini2014}, which can provide $>$1\,km light-matter interaction pathlengths. However, the increased detection sensitivity of cavity-based techniques applies equally to all resonant transitions of every molecular species inside the interaction volume. The lack of sufficient detection selectivity is problematic in the ``molecular-fingerprint'' mid-infrared (mid-IR) range. Because of high density of strongly overlapping transitions, spectroscopic detection and assignments of weak mid-IR signals can be prohibitively difficult with conventional cavity-enhanced techniques.

The development of optical detection for the rare radiocarbon dioxide molecule ($^{14}$CO$_2$), with $\sim$1200 parts-per-quadrillion (10$^{15}$, ppq) $^{14}$C/C natural abundance, exemplifies this need for a spectroscopic technique that simultaneously achieves high detection $sensitivity$, $selectivity$, and quantitation $accuracy$~\cite{mccartt2022room,jiang2021two}. Traditionally measured by accelerator mass spectrometry (AMS)~\cite{bennett1977radiocarbon,nelson1977carbon}, the $^{14}$C tracer (half-life of 5730$\pm$40 years)~\cite{godwin1962half} has been used in a wide range of applications, such as archaeological dating~\cite{taylor2016radiocarbon}, bio-medicine development~\cite{turteltaub2000bioanalytical,wong2023design}, earth carbon-cycle studies~\cite{heaton2021radiocarbon}, and monitoring of fossil-fuel-CO$_2$ emission~\cite{sargent2018anthropogenic,miller2020large,basu2020estimating}. $In$ $situ$ field measurements of $^{14}$C are not possible with AMS, which utilizes a room-size, mega-volt accelerator to filter the interfering molecular isobars of $^{14}$C (e.g., $^{13}$CH). Even for laboratory measurements, the investment and operational cost of AMS (multiple million dollar in equipment and staff) are too high for many applications.

Mid-IR detection of $^{14}$CO$_2$, by measuring its $\nu_3$-band ro-vibrational transitions, has been proposed as a cheaper and potentially field-deployable $^{14}$C sensing technique~\cite{mccartt2022room,jiang2021two,galli2011molecular,galli2016spectroscopic,mccartt2016quantifying,fleisher2017optical,genoud2015radiocarbon,terabayashi2020mid,kratochwil2018nanotracing}. Quantifying fossil-fuel-CO$_2$ emission based on measurements of the total atmospheric CO$_2$ content is subject to the $large$ and $highly$ $variable$ CO$_2$ emissions from the bio-sphere~\cite{miller2020large,basu2020estimating,sargent2018anthropogenic}. Combustion of fossil fuels, which are depleted of $^{14}$C, leads to location- and time-dependent decrease in the atmospheric $^{14}$CO$_2$:CO$_2$ ratio, with the measured dip typically $<$100\,ppq (i.e., $\lessapprox$10$\%$ of natural $^{14}$CO$_2$ concentration) in a mega-city~\cite{miller2020large}. This signature dip is an $unambiguous$ $gold$-$standard$ tracer for fossil-fuel-CO$_2$. Large-scale and year-long measurement campaigns of atmospheric $^{14}$CO$_2$ have only been occasionally implemented in a very few locations in the world, because of the high costs of AMS measurements~\cite{miller2020large,basu2020estimating}. A compact field-deployable $^{14}$CO$_2$ sensor will provide accurate, permanent, and low-latency monitoring of fossil-fuel-CO$_2$ emission, and thereby facilitates evaluating the efficacy of various carbon reduction programs~\cite{mitchell2022multi}. 

Optical detection of $^{14}$CO$_2$ is challenging at the concentration ($\lesssim$\,natural abundance) and accuracy level (1-100\,ppq $^{14}$C/C) required for many of the aforementioned applications. Mid-IR detection of $^{14}$CO$_2$ in room-temperature samples with better than 10-ppq sensitivity and accuracy, demonstrated in this work with two-color cavity ringdown spectroscopy (2C-CRDS), pushes the limit of mid-IR laser absorption techniques in $sensitivity$, $selectivity$, and $accuracy$. Our current 2C-CRDS setup allows measurements of a minimum absorption coefficient ($k$) of $4\times10^{-13}$\,cm$^{-1}$ from  $^{14}$CO$_2$ in the presence of strong one-color (1C) hot-band absorption signals from other CO$_2$ isotopologues, with the background $k$ typically $>$$10^{-7}$ cm$^{-1}$ (4.55 $\mu$m, 20 torr, 300\,K)~\cite{gordon2022hitran2020}. This background/signal ratio ($>$10$^5$) is too large for $^{14}$CO$_2$ detection by other cavity-enhanced techniques based on single-photon absorption. To mitigate severe spectral overlap, gas-cooling to 170\,K has been necessary for previous 10-ppq level measurements of $^{14}$CO$_2$, which was achieved by a 1C variant of CRDS, the saturated-absorption cavity ringdown (SCAR) technique~\cite{galli2016spectroscopic,delli2021biogenic}. The gas-cooling requirement for SCAR increases instrumental complexity and size, and is not ideal for field-work applications. 

\begin{figure*}
\centering
\includegraphics[width=6 in]{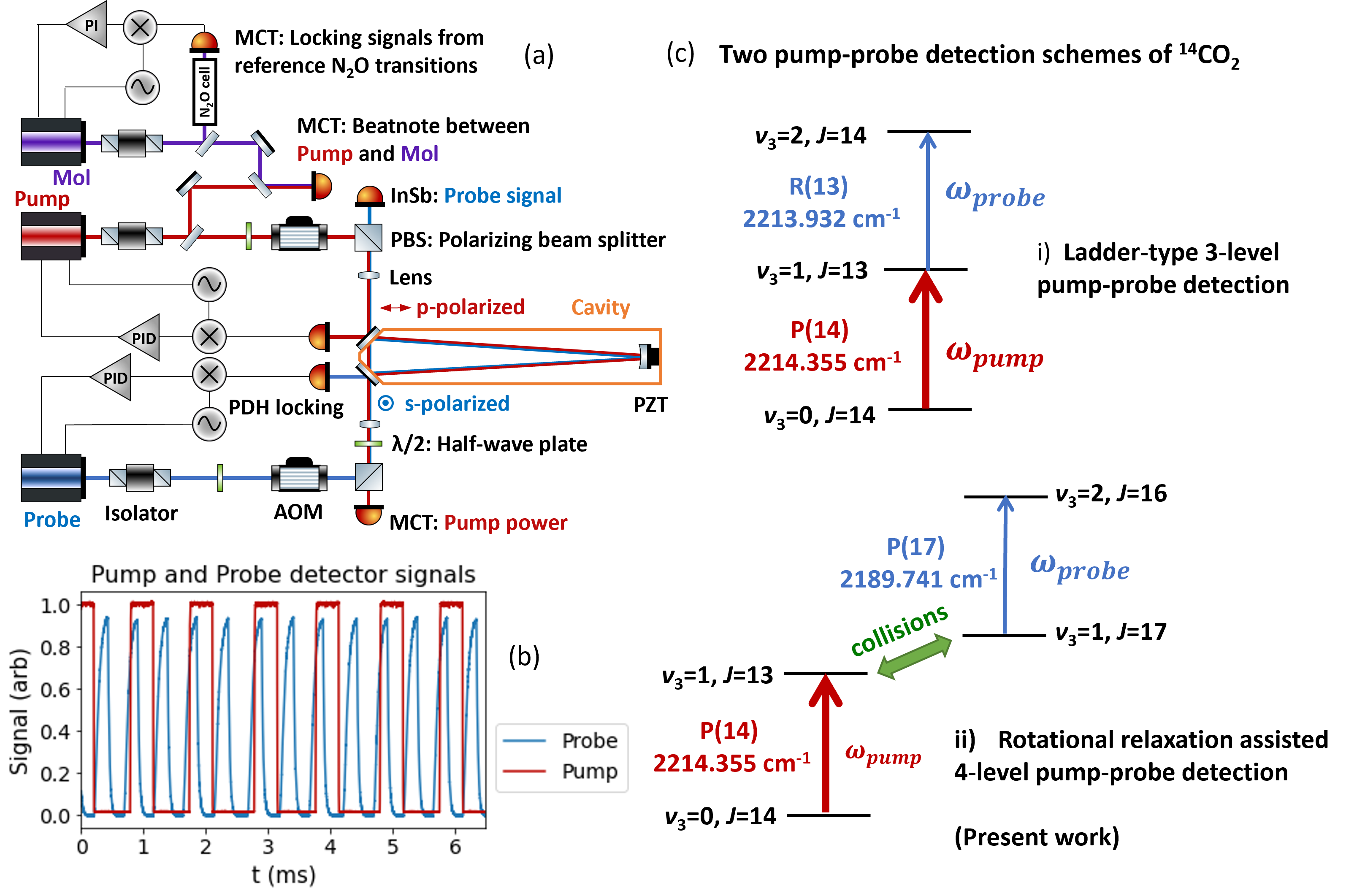}
\caption{2C-CRDS experimental schemes. (a) Experimental schematic. The counter-propagating pump and probe beams are coupled, respectively, to a $p$- (finesse=5300) and $s$-polarization (finesse=67700) mode of the three-mirror cavity. See Methods and SI Appendix, Section S1.2 for further details on the 2C-CRDS technique and the detection system.  (b) Time traces of the pump and probe signals. (c) Diagrams that show two pump-probe detection schemes with 2C-CRDS, scheme (i) for our previous work~\cite{mccartt2022room,jiang2021two} and scheme (ii) for the present work.}
\label{fig:schemes}
\end{figure*}

The built-in baseline compensation capability of 2C-CRDS detection leads to its significantly enhanced sensitivity, selectivity, and quantitation accuracy relative to conventional CRDS methods~\cite{mccartt2022room,jiang2021two}. In our experiment (Fig.~\ref{fig:schemes}a), the outputs from two quantum cascade lasers (QCL) excite a pair of $\nu_3=1\leftarrow0$ (pump) and $\nu_3=2\leftarrow1$ (probe) ro-vibrational transitions of $^{14}$CO$_2$ inside a three-mirror, traveling-wave cavity. With the pump radiation switched off during alternative probe ringdown events (Fig.~\ref{fig:schemes}b), the net 2C signals are immune to the drift of the cavity ringdown rates and signals from one-photon molecular transitions. 

The 2C-CRDS method has been previously applied by our group to achieve the first-ever room-temperature optical detection of $^{14}$CO$_2$ below its natural abundance, with measurement accuracy of $\sim$100 ppq ($\sim$8$\%$ natural abundance)~\cite{mccartt2022room}. A three-level excitation (scheme i in Fig.~\ref{fig:schemes}c) is used to quantify the $^{14}$CO$_2$ concentrations of several combusted $^{14}$C ``standard'' samples. The observed 2C-CRDS spectra are free of interference from one-photon hot-band transitions of other CO$_2$ isotopologues that lead to $>$10000\,s$^{-1}$ ringdown rate loss~\cite{mccartt2022room}. However, small background 2C signals ($\sim$6.5\,s$^{-1}$ ringdown rate loss) are observed near the $\nu_3=2\leftarrow 1$, R(13) probe transition of $^{14}$CO$_2$. Collisional excitation of vibrationally excited levels of other CO$_2$ isotopologues, which are inadvertently populated by the strong intra-cavity pump radiation, are believed to be the cause of this 2C background. 

These collision-induced signals are significantly more sensitive to changes in the experimental conditions than the 2C signals from $^{14}$CO$_2$ (SI Appendix, Section~S3.1). Unlike the pump-power saturated $^{14}$CO$_2$ signals, the background signals have a linear dependence on the pump power. In addition, the background signals are sensitive to small fluctuations of the gas temperature ($\sim$1$\%$ signal variation from a 0.1$^{\circ}$C temperature change), because of the involvement of hot-band pump excitation of CO$_2$ levels in the 5000\,cm$^{-1}$ energy region. Thanks to the relatively small magnitude of this background, equivalent to $^{14}$CO$_2$ signals at 1.5$\times$ natural abundance (1800\,ppq), 2C-CRDS detection of $^{14}$CO$_2$ was still feasible below its natural abundance in our previous work, given the moderately stable gas temperature ($\sim$0.1$^{\circ}$C variation) and pump power ($<$5$\%$ variation) during the experiments. However, to achieve ppq-level detection accuracy, significant further background reduction is imperative.

After achieving $\sim$$10\times$ reduction in the background 2C signal (guided by a collision model) and $25\times$ improvement in the detection signal-to-noise ratios, we have accomplished, by 100-s averaging at the maximum of the $^{14}$CO$_2$ 2C peak, optical detection of $room$-$temperature$ $^{14}$CO$_2$ with 7-ppq accuracy. The accuracy further improves to 4 ppq after fitting the 2C-CRDS spectra (20-30 min data acquisition). This record sub-10-ppq measurement performance (the most $sensitive$ and $accurate$ in the mid-IR) has been reproducibly demonstrated with several rounds of measurements of combusted $^{14}$C standards and low-$^{14}$C-content bio-fuel samples (10-80\,ppq). The high sensitivity, high selectivity, and high accuracy measurement capabilities of the 2C-CRDS technique will have significant impact on analytical trace measurements and fundamental gas-phase chemical physics studies, which are discussed at the end of this paper.

\section*{Room-temperature \lowercase{ppq}-level measurements of $^{14}$CO$_2$}

\begin{figure*}
\centering
\includegraphics[width=6.3 in]{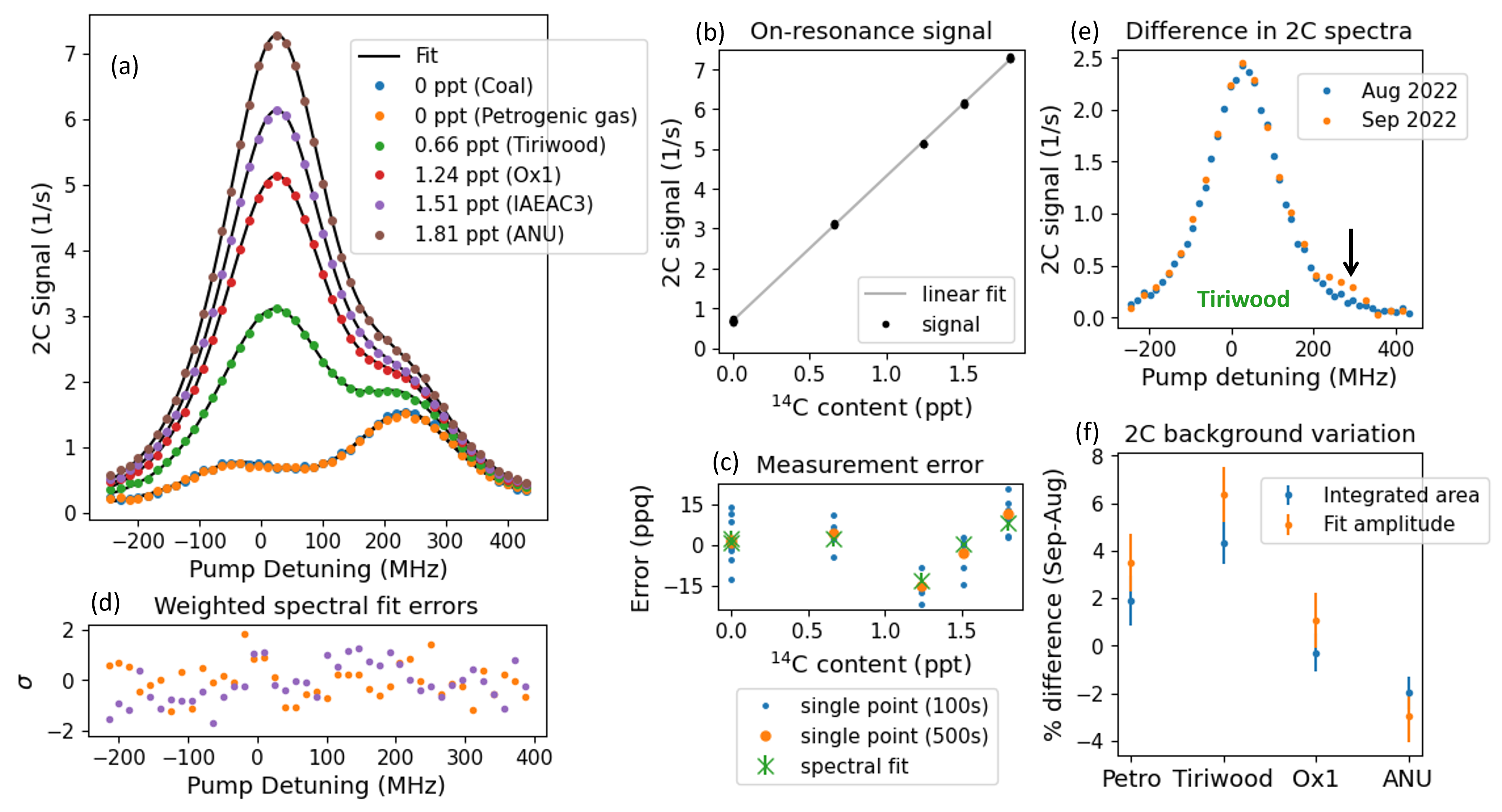}
\caption{2C-CRDS measurements of combusted $^{14}$C standard samples (20.1\,torr) (SI Appendix, Section S2). (a) 2C-CRDS spectra (100-s averaging per data point), and their spectral fit models. Note that ``ppt'' stands for part-per-trillion (10$^{12}$). (b) Comparison of the on-resonance, fixed-frequency 2C-CRDS signals with the sample $^{14}$C contents. (c) Measurement errors for the sample $^{14}$CO$_2$ concentrations, based on the residuals of the linear fit in panel (b). (d) Weighted spectral-fit errors for two 2C-CRDS spectra in panel (a). See SI Appendix, Section S1.1 for the statistical weights ($\sigma$) used in the fit. (e) Differences in the 2C-CRDS spectra of two ``Tiriwood'' samples from the ``Aug 2022'' and ``Sep 2022'' measurements (see Table~\ref{tab:fit}). The arrow highlights the extra background signals present in the ``Sep 2022'' spectrum. (f) Variations of the background 2C signals from four overlapping sample types from ``Aug 2022'' and ``Sep 2022''. The ``Integrated area'' values are obtained by numerical integration of the observed spectra, and the ``Fit amplitude'' values for the background 2C signals are derived from the spectral fit.}
\label{fig:standards}
\end{figure*}

The use of a three-level pump-probe scheme with a common intermediate level, such as our original $\nu_3=1\leftarrow 0$, P(14) pump and $\nu_3=2\leftarrow 1$, R(13) probe combination, is not necessary for $^{14}$CO$_2$ detection in a static-gas cavity at $\sim$20 torr. The vibrational relaxation rate of the $\nu_3=1$ state of $^{14}$CO$_2$ ($\sim$30\,ms$^{-1}$torr$^{-1}$, determined from a pump-probe delay experiment similar to that on N$_2$O)~\cite{jiang2021two} is significantly slower than its rotational relaxation rate (on the order of 0.1\,ns$^{-1}$torr$^{-1}$)~\cite{yardley2012introduction}. As a result of facile rotational relaxation and negligible diffusion loss at 20\,torr ($\gg$10$\times$ slower than the $^{14}$CO$_2$ vibrational relaxation loss), a population distribution that resembles a thermal distribution at 300\,K exists among the $\nu_3=1$ rotational levels under continuous pump excitation during the ``pump on'' cycle, even though only one $J$-level in $\nu_3=1$ is directly populated by the pump.

A rotational-relaxation-assisted, four-level detection scheme of $^{14}$CO$_2$, $\nu_3=1\leftarrow 0$, P(14) pump~\cite{galli2011v3} and $\nu_3=2\leftarrow 1$, P(17) probe~\cite{zak2017room,huang2017ames}, is used in the measurements presented here (scheme ii in Fig.~\ref{fig:schemes}c). The background 2C signal is $\sim$10$\times$ smaller at the probe resonance frequency for this P(14)-P(17) combination than the original P(14)-R(13) scheme. 

\begin{figure}
\centering
\includegraphics[width=4.2 in]{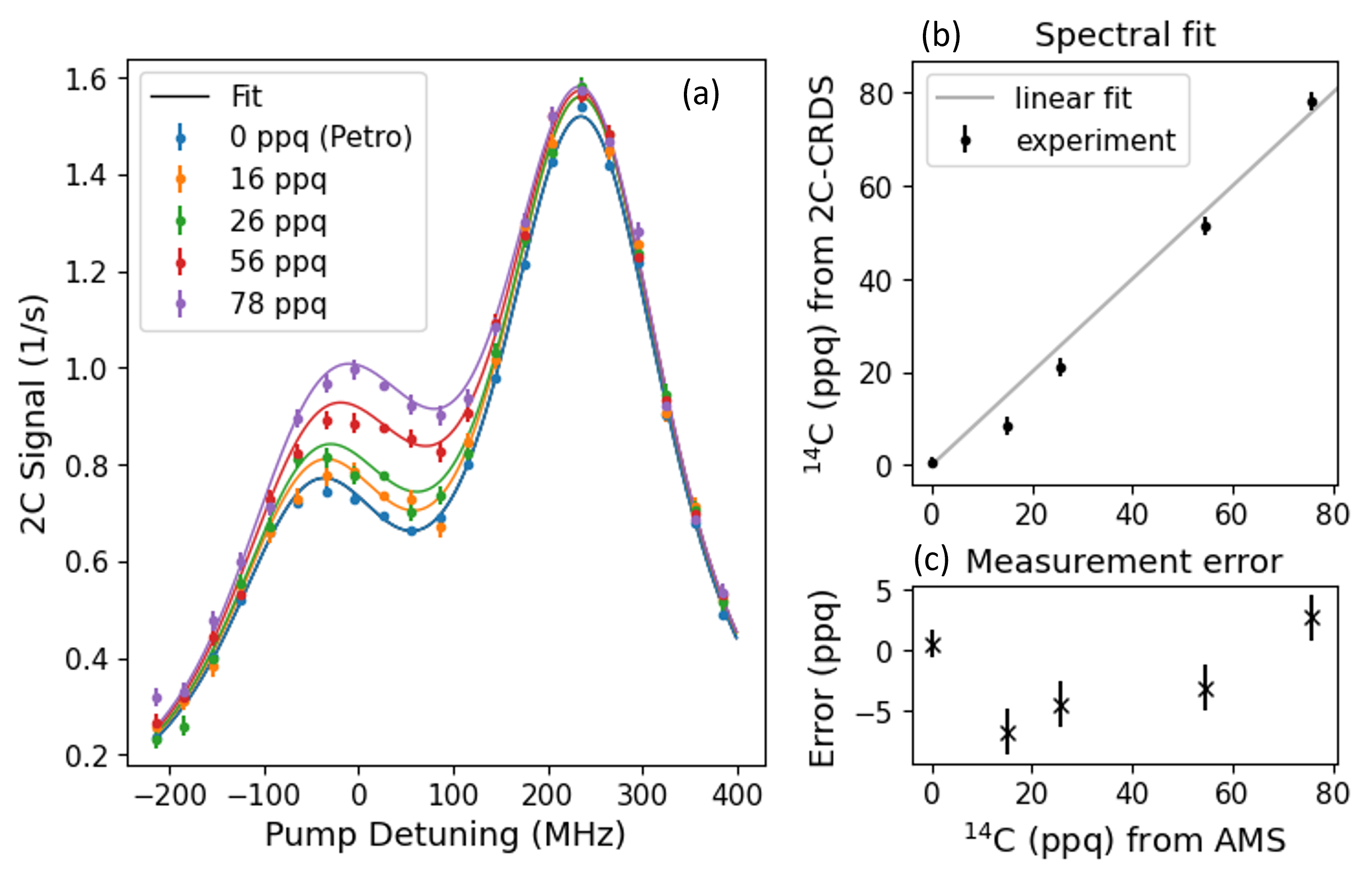}
\caption{2C-CRDS measurements of combusted bio-fuel samples (20.1\,torr) (SI Appendix, Section S2). (a) 2C-CRDS spectra (60-s averaging per data point), and their spectral fit models. (b) Comparison of the sample $^{14}$C content determined by 2C-CRDS (spectral fit) and AMS. (c) Measurement errors for the sample $^{14}$C content, based on the deviations from the line in panel (b). The errorbars indicate the standard errors of the spectral fit for the amplitudes of the $^{14}$CO$_2$ signal in the observed 2C spectra. The $^{14}$C contents for the four bio-fuel samples are calibrated based on the 2C-CRDS measurements of ``Petrogenic gas'' and combusted ``ANU'' (1.81 ppt $^{14}$C/C) samples. These two types of samples were measured daily with these four bio-fuel samples.}
\label{fig:biofuels}
\end{figure}

The 2C-CRDS technique allows, with very high signal-to-noise ratios, differentiation of six combusted $^{14}$C standard samples, for which the $^{14}$C content ranges from 0 to 1.5$\times$ natural abundance (Fig.~\ref{fig:standards}a). Similarly, despite the very low $^{14}$C content (10-80 ppq), 2C-CRDS measurements of the four bio-fuel samples yield different signal levels at the $^{14}$CO$_2$ transition region with only 60-s averaging per data point (Fig.~\ref{fig:biofuels}a). The magnitude of the collision-induced background at the maximum of the $^{14}$CO$_2$ 2C peak (determined from the ``Coal'' and ``Petrogenic gas'' samples) is equivalent to that from 210\,ppq of $^{14}$CO$_2$. Considering that the 2C baseline is essentially flat from a $^{12}$C-enriched and $^{14}$C-depleted CO$_2$ sample, collision-induced 2C transitions of at least one of the six $^{13}$C isotopologues of CO$_2$ must be responsible for the observed background signals in Figs.~\ref{fig:standards}a and \ref{fig:biofuels}a. This observation agrees with the results of our model for the collision-induced processes relevant to 2C-CRDS detection, which suggests hot-band pump excitation of $^{13}$C$^{16}$O$_2$ as the cause of the remaining background.

For each of the six combusted $^{14}$C samples in Fig.~\ref{fig:standards}a, the signal at the maximum of the $^{14}$CO$_2$ 2C transition is measured five times in 1.5 hours, each for a duration of 100\,s. These 2C signals at fixed pump-probe frequencies scale linearly with the $^{14}$C content of the corresponding samples (Fig.~\ref{fig:standards}b). Residuals from a linear fit to the 100-s measurements (Fig.~\ref{fig:standards}c) have a mean absolute error equivalent to 0.7\,$\%$ of the $^{14}$CO$_2$ natural abundance (8.4\,ppq). The measurement accuracy improves to 6.1\,ppq after averaging the five 100-s measurements of each sample. Because of effective background compensation, the 2C signals measured with the fixed-frequency approach are highly repeatable, with month-to-month stability for their $absolute$ intensities at the 10-ppq level. However, given that the amount of improvement in the measurement accuracy (8.4\,ppq$\rightarrow$6.1\,ppq) is smaller than expected based on the amount of increase in averaging time (100\,s$\rightarrow$500\,s), the fixed-frequency measurements must have suffered from small systematic errors. Certain types of errors, such as variability in the sample $^{13}$C content (1-2$\%$ typical) and variations of the background 2C signal due to changes in the experimental conditions (Figs.~\ref{fig:standards}e and \ref{fig:standards}f), can be compensated by spectral fitting (SI Appendix, Section S1.1). For all four trial measurements of combusted $^{14}$C samples, the spectral fit approach consistently yields improved measurement accuracy (4.0\,ppq) compared to the fixed-frequency method (Table~\ref{tab:fit}). 

\begin{table*}[b]
\caption{Summary of 2C-CRDS measurements of combusted $^{14}$C standards and bio-fuel samples (20.1\,torr).}
\begin{center}
\begin{tabular}{|@{\hspace{8pt}} c @{\hspace{8pt}} @{\hspace{6pt}} c @{\hspace{6pt}}  @{\hspace{6pt}} c   @{\hspace{6pt}}c  | @{\hspace{6pt}} c  @{\hspace{6pt}}  | @{\hspace{6pt}} c  @{\hspace{6pt}} | c   @{\hspace{6pt}} c @{\hspace{6pt}}  @{\hspace{6pt}} | c   @{\hspace{6pt}}c  | @{\hspace{6pt}} c @{\hspace{6pt}} | @{\hspace{6pt}} c @{\hspace{6pt}}}

\hline
Samples & Measurement & $^{14}$C range & & 100\,s fixed  & 500\,s fixed & & Spectral fit\\
 & periods & (ppq)   & &(ppq) & (ppq) && (ppq)\\
\hline
6 standards & Aug. 2022 (3 days) & 0-1800 & &	8.4 & 6.1 && 4.5\\
4 standards & Sep. 2022 (1 day) & 0-1800 & &	6.6 & 3.5 &&0.9\\
7 standards & Nov. 2022 (3 days) & 0-1800 & &	8.5& 7.9 &&6.5\\
4 bio-fuels & Dec. 2022 (4 days) & 10-80 & &	6.1 & 6.1 && 4.2\\
\hline
 &  & $Average$ & &6.9 & 5.1 &&4.0\\
\hline
\end{tabular}
\end{center}
\label{tab:fit}
\end{table*}

Prior to our current results, the SCAR technique achieved the most sensitive measurements in the mid-IR. By utilizing the high intra-cavity power from a cavity-locked probe, SCAR allows simultaneous measurements of the empty-cavity ringdown rate and the gas-induced absorption. With baseline compensation and 2-hour signal averaging, SCAR achieved 10-ppq measurement of $^{14}$CO$_2$ at 170\,K~\cite{galli2016spectroscopic,delli2021biogenic}. Room-temperature detection of $^{14}$CO$_2$ is not possible with SCAR, even above the natural abundance, because of the overwhelmingly large background 1C signals from other CO$_2$ isotopologues.

\section*{Current optical detection sensitivity}

In our initial 2C-CRDS measurements~\cite{mccartt2022room,jiang2021two}, the beginning of each probe ringdown transient was contaminated by random oscillations with amplitudes much larger than the detector noise~\cite{lehmann2009optimal}. We have shown that the shot-to-shot ringdown rate fluctuations ($\sigma_{sts}$) of our detection system can be reduced by 25$\times$, after a small current is applied to the probe laser current driver concurrent with the trigger for the probe AOM. This fast current injection, which is achieved by temporarily setting an incorrect ``zero'' level for the locking servo of the probe laser, detunes the probe laser frequency from the original cavity resonance. The combination of this laser-frequency-jump and the usual AOM-controlled beam shut-off leads to a ``cleaner'' initiation of the ringdown events than the use of an AOM alone~\cite{balslev2011application,huang2009noise}. The extra noises in the original setup are most likely caused by interference between the ringdown signal and the leaked probe radiation through the AOM due to its finite light extinction ratio~\cite{huang2009noise}. A $\sigma_{sts}$ value of 5\,s$^{-1}$ for the 2C signal with our current detection system (i.e., a shot-to-shot $k$ value of $1.7\times10^{-10}$ cm$^{-1}$) is only 25$\%$ higher than the $single$-shot noise-equivalent signal from $^{14}$CO$_2$ at natural abundance. The short-term detection sensitivity reaches $1.7\times10^{-13}$\,cm$^{-1}$ after 23-minute averaging, based on Allan deviation analysis of 2C-CRDS signals from multiple $^{14}$CO$_2$ samples (SI Appendix, Fig.~S1). To our knowledge, this ``ultimate sensitivity'' level, equivalent to $^{14}$CO$_2$ signals at 1.4 ppq concentration, is better than any previous optical measurements in the mid-IR.

\section*{Collision-induced 2C background}

\begin{figure}
\centering
\includegraphics[width=5.25 in]{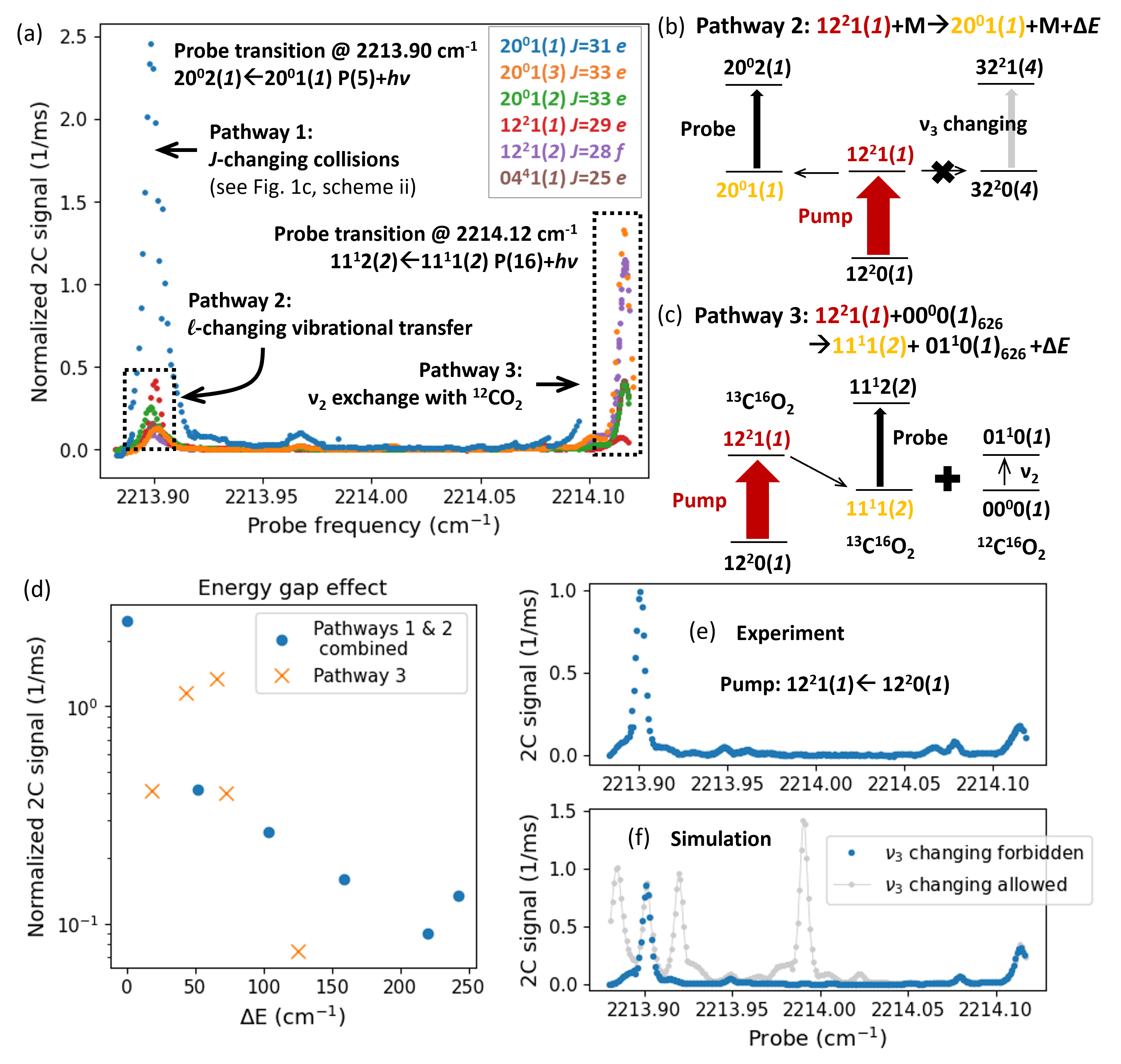}
\caption{Collision-induced 2C signals following pump excitation of selected hot-band transitions of $^{13}$CO$_2$. (a) Overview of six 2C spectra. The amplitude of each spectrum is normalized based on the spectral line intensity of the corresponding pump transition~\cite{huang2022ames} and the measured pump power during the experiment (assuming a linear pump power dependence). In the $v_1v_2^lv_3(P)$ notation, $v_i$'s are the nominal quantum numbers in the three corresponding vibrational modes, $l$ is the vibrational angular momentum quantum number, and $P$ indicates the energy rank for vibrational states that belong to the same Fermi interaction polyad (i.e., with the same values of 2$v_1$+$v_2$, $v_3$, and $l$). (b) Level diagram for collision Pathway 2 and the forbidden $\nu_3$-changing pathway from the pump-populated $12^21(1)$ state. The species M in panel (b) is predominantly the $^{12}$C$^{16}$O$_2$ molecule. (c) Level diagram for collision Pathway 3 from the pump-populated $12^21(1)$ state. (d) 2C peak intensities in panel (a) as a function of the energy gap for the corresponding collisional mechanisms. (e) Experimental spectrum following pump excitation into the $12^21(1)$, J=29\,$e$ level. (f) Simulated spectrum for (e), with two assumptions regarding changes for the $\nu_3$ quantum number.}
\label{fig:collisions}
\end{figure}

Adoption of the current P(14)-P(17) detection scheme for $^{14}$CO$_2$ is guided by the results of a series of pump-probe experiments to study collision-induced background 2C signals from other CO$_2$ isotopologues. In each of those experiments, the pump directly populates a rotational level in one of the following six nearly-degenerate vibrational states of $^{13}$CO$_2$ in the 5100-5300 cm$^{-1}$ energy region: $04^41(1)$, $12^21(1)$, $12^21(2)$, $20^01(1)$, $20^01(2)$, and $20^01(3)$ (see Fig.~\ref{fig:collisions} caption for the $v_1v_2^lv_3(P)$ notation used for vibrational assignments). Probe signals from two $^{13}$CO$_2$ transitions, $20^02(1)$$\leftarrow$$20^01(1)$ P(5$e$) (2213.90\,cm$^{-1}$) and $11^12(1)$$\leftarrow$$11^11(2)$ P(16$e$) (2214.12\,cm$^{-1}$), are observed in all six experiments. 

Three dominant collisional relaxation pathways (Fig.~\ref{fig:collisions}a) and propensity rules on the $\nu_3$ and $l$ quantum numbers can be identified from these experiments. Efficient $J$-changing collisions (Pathway 1) occur within each vibrational state. $l$-changing population transfer is observed from five of the pump-populated $^{13}$CO$_2$ vibrational states into the nearly-degenerate $20^01(1)$ (Pathway 2, see Fig.~\ref{fig:collisions}b). The pump-populated level also exchanges quanta of $\nu_2$ with the background $^{12}$CO$_2$ bath (Pathway 3, see Fig.~\ref{fig:collisions}c) with a small energy change ($\Delta E$). For Pathway 2 and 3, the intensities of the 2C peaks, in general, decrease by $\sim$$10\times$ for every 100-200\,cm$^{-1}$ increase in $\Delta E$ (Fig.~\ref{fig:collisions}d). Note that the $\nu_3$ quantum number is $conserved$ in all three observed pathways in Fig.~\ref{fig:collisions}a. The $\nu_3$-$changing$ vibrational energy transfer between nearly-degenerate states, e.g., $12^21(1)$$\rightarrow$$32^20(4)$ with $\Delta E\sim20$\,cm$^{-1}$ (see Fig.~\ref{fig:collisions}b), is significantly less efficient than these three pathways. 

The identification of three dominant collisional processes, together with the use of the ``energy-gap law'' and quantum-number propensity rules~\cite{yardley2012introduction}, allows us to model the collision-induced 2C spectra in Fig.~\ref{fig:collisions}a (SI Appendix, Sections S3.2 and S3.3). According to our simulation for pump excitation into $12^21(1)$, J=29$e$ (Fig.~\ref{fig:collisions}f), additional collision-induced 2C peaks would have been observed in the corresponding probe spectrum (Fig.~\ref{fig:collisions}e), if the $\nu_3$-changing population transfer were allowed. The absence of these additional features strongly supports our proposed propensity rule on the conservation of the $\nu_3$ quantum number during 2C-CRDS measurements of CO$_2$.

We have simulated the background 2C signals at various pump-probe combinations for $^{14}$CO$_2$ detection, based on the fit model derived from Fig.~\ref{fig:collisions} (SI Appendix, Section S3.3). The use of collision-assisted four-level detection significantly improves the likelihood of finding a pump-probe combination with reduced 2C background. With the same $\nu_3=1\leftarrow0$, P(14) pump transition as in the original experiment, background signals of $<$10\,s$^{-1}$ are predicted at the resonance frequencies of every $\nu_3=2\leftarrow1$, P-branch transition from P(31) to P(7). We measure the background signals with six of these P-branch probes, P(27) to P(17), that fall within the tuning range of the available QCL in our laboratory. For four of these P-branch transitions, the background signal is considerably smaller than that from the original R(13) probe (6.5\,s$^{-1}$). The current P(17) probe yields the smallest observed background signal. Work is ongoing to investigate other pump-probe combinations for further reduction of background 2C signals.

\section*{Implications and outlook}

By monitoring the baseline fluctuations and background 1C absorption during alternative probe ringdown events, the 2C-CRDS method significantly enhances laser spectroscopic detection in $sensitivity$, $selectivity$, and $quantitative$ $accuracy$. In combination with recent advances in laser radiation sources, detectors, and mirror coatings in the mid-IR, the technique will greatly enhance the utility of high-resolution mid-IR detection for analytical and spectroscopic studies.

In addition to atmospheric $^{14}$CO$_2$ measurements, $in$ $situ$ mid-IR detection of trace reactive radical molecules in the atmosphere, such as OH, HO$_2$, and NO$_3$, can be achieved with the 2C-CRDS method. Field measurements of atmospheric radicals  provide valuable experimental inputs for evaluating different models of oxidation chemistry in the earth's troposphere~\cite{heard2003measurement,stone2012tropospheric}. Among these radicals, detection of OH, often referred to as the ``detergent of the atmosphere'' for removing CH$_4$ and other harmful gases (e.g., CO and volatile organic compounds), is particularly challenging because of its very low steady-state concentration, in the range of 40-400 ppq by volume (ppqv) during daytime. A mid-IR instrument will be lighter, more compact, and cheaper than the existing UV spectrometers for direct OH detection, which is based on either the fluorescence assay by gas expansion (FAGE) technique or multi-pass differential optical absorption spectroscopy (DOAS) to measure the $A^2\Sigma^+-X^2\Pi_i$ transitions. Large and heavy high-throughput vacuum pumps, necessary for FAGE to avoid detection of the OH artifacts generated by the probe UV pulses~\cite{heard2003measurement,stone2012tropospheric}, would not be needed for mid-IR detection. While OH artifacts do not affect DOAS, which uses a low-intensity probe, the DOAS technique is not compatible with airborne measurements because of its large footprint ($>$1\,km total pathlength with 10-40\,m mirror separation)~\cite{heard2003measurement}. 

Mid-IR detection of atmospheric OH encounters challenges similar to those for $^{14}$CO$_2$. One-photon detection of OH (using $X^2\Pi_{3/2}$, $v=1-0$ transitions near 3570\,cm$^{-1}$) is not possible because of spectral overlap with the $\nu_3$-band transitions of water (3750\,cm$^{-1}$). An intra-cavity pump-probe scheme, e.g., $X^2\Pi_{3/2}$, $v=1-0$, P(2.5e/f) and $v=2-1$, Q(1.5e/f), would allow accurate 2C-CRDS measurements of atmospheric OH concentrations without significant interference from nearby water transitions. Even though the transition dipole moments for OH ro-vibrational excitation are $\sim$10$\times$ smaller than those for the $\nu_3$-band transitions of CO$_2$~\cite{gordon2022hitran2020}, detection sensitivity of $\sim$50\,ppqv OH could be achieved based on the sensitivity of our current setup. For detection of radical species, background 2C signals from closed-shell molecules (e.g., water and CO$_2$) can be further filtered by taking advantage of the much larger Zeeman effects of radicals. The measurement sensitivity, selectivity, and accuracy of radical species will all be significantly enhanced with the incorporation of AC Zeeman modulation~\cite{zhao2011sensitive,pfeiffer1981sensitive,lewicki2009ultrasensitive} in the 2C-CRDS method. 

The 2C-CRDS technique provides a new mid-IR detection scheme for spectroscopic and chemical dynamics studies. The technique will enable probing chemical species at concentrations, internal energies, and conformations that are not easily accessible with other methods. The incorporation of a mid-IR frequency comb~\cite{schliesser2012mid} as the probe is a particularly attractive direction for the 2C-CRDS technique. With single-frequency pump and broadband probe, the technique will enable high-sensitivity, high-selectivity, and $multiplexed$ investigation of the molecular level structure in the high internal energy region of the electronic ground state of many molecules. Classic, fundamental problems in chemical physics, such as intramolecular vibrational energy redistribution, isomerization, and bond dissociation will be systemically explored in a highly-$sensitive$ and level-$specific$ manner. In combination with a Chen-type hyperthermal nozzle~\cite{kohn1992flash}, which produces a vibrationally-hot but rotationally-cold population distribution~\cite{changala2015probing}, the 2C-CRDS method and its cavity-enhanced 2C variants complement (and, in certain applications, exceed) the capabilities of the widely-used stimulated emission pumping technique for studying molecular dynamics in the ground electronic state~\cite{hamilton1986stimulated,dai1995molecular}, in particular for molecules with only short-lived ($<$1\,ns) electronically excited states.

\section*{Methods}

\subsection*{2C-CRDS detection}
\label{sec:2ccrds}

With a 67-cm round trip, the free spectral range (FSR) of the three-mirror cavity is 443.3\,MHz. The $s$- and $p$-mode cavity resonances are interleaved with a spacing of  $\sim$$\frac{\textup{FSR}}{2}$, because of a net $\sim$$\pi$-phase shift between the two polarizations upon reflection inside the three-mirror cavity~\cite{saraf2007high}. Unlike the free-space pump-probe experiments, the pump laser frequency cannot be set at a fixed value in an intra-cavity excitation scheme. A change in the cavity FSR value leads to a shift in both the pump and probe laser frequencies~\cite{mccartt2022room,jiang2021two}. As a result, in general, the pump ($p$-polarized) and probe ($s$-polarized) frequencies will not be simultaneously on-resonance with their respective target molecular transitions in our 2C-CRDS experiments. For $^{14}$CO$_2$ detection reported in this work, the pump radiation is coupled to a cavity $p$-mode resonance that lies closest to the resonance frequency of a target $^{14}$CO$_2$ pump transition. Under this experimental scheme, 2C signals from $^{14}$CO$_2$ are observed with a maximum absolute pump detuning frequency of $\frac{\textup{FSR}}{2}$ (221.65\,MHz), regardless of the choice of the pump and probe transitions. Because of the strong cavity-enhanced pump power (20\,W) that leads to significant power saturation and broadening ($\sim$300\,MHz at full-width half-maximum) of the pump transition, the observed 2C signals are minimally affected by the absence of double-resonance excitation condition.

The strong and sustained intra-cavity pump power is achieved by stabilizing the pump laser frequency to a cavity resonance with the Pound-Drever-Hall (PDH) method~\cite{drever1983laser}. The output frequency from a third laser (``Mol'' in Fig.~\ref{fig:schemes}a) is locked to the center of an N$_2$O transition. The beatnote between this third laser and the pump is used to stabilize the cavity length, and to calibrate both the pump and probe laser frequencies. Further details on our 2C-CRDS detection system are provided in SI Appendix, Section S1.2.

The resonance frequencies of the $\nu_3=1\leftarrow0$ P(14) (pump) and $\nu_3=2\leftarrow1$ P(17) (probe) transitions of $^{14}$CO$_2$~\cite{galli2011v3,zak2017room,huang2017ames} are separated by nearly an exact odd integer multiple of the frequency spacing between neighboring $p$- and $s$-mode cavity resonances of our cavity (i.e., $\sim$$3329\times\frac{\textup{FSR}}{2}$). As a result, the $^{14}$CO$_2$ 2C transition from the current P(14)-P(17) pump-probe scheme occurs at the near double-resonance excitation condition, with a frequency detuning of $\sim$30\,MHz and $\sim$0\,MHz, respectively, for the pump and probe lasers. In addition, the $J=17$ level is near the maximum of the room-temperature rotational distribution of CO$_2$. The observed $^{14}$CO$_2$ signals in our current experiments are thus nearly maximized for intra-cavity pump-probe detection of the molecule.


As in our previous work~\cite{mccartt2022room}, 2C-CRDS measurements of the CO$_2$ gas samples are all taken at 20.1 torr. At this measurement pressure, the probe transition of $^{14}$CO$_2$ is weakly saturated, considering that the magnitude of the $^{14}$CO$_2$ 2C signals depends on the starting voltage ($V_0$) of the ringdown fit, e.g., a 50$\%$ decrease of $V_0$ leads to $\sim$30$\%$ increase in the 2C signals. Note that, in the strongly-saturated regime, the gas-absorption-induced ringdown rate ($\gamma_{gas}$) approaches zero, while $\gamma_{gas}$ is independent of $V_0$ in the non-saturated limit. The background 2C signals at the $^{14}$CO$_2$ pump-probe transition region relevant to this work is significantly less saturated than the $^{14}$CO$_2$ probe transition. While the degree of saturation of the $^{14}$CO$_2$ probe transition can be reduced at higher gas pressures, collision-induced homogeneous broadening will lead to an increase in the background 2C signal level. We are currently working on optimizing various experimental conditions, such as the gas pressure and selection of pump-probe transitions (for further reduction of the background 2C signal), to improve the sensitivity and accuracy of 2C-CRDS detection of $^{14}$CO$_2$.

\printbibliography



\section*{Acknowledgments}
We thank Professor Robert W. Field (MIT), Stephan L. Coy (MIT), and Professor Kevin K. Lehmann (UVA) for insightful and encouraging comments on the manuscript, and Bruce Buchholz, Kari Finstad, Esther Ubick, and Ted Ognibene (LLNL) for their assistance with the sample preparations.

{\bf Funding:} Research reported in this publication was supported by the National Institute Of General Medical Sciences of the National Institutes of Health (Award Number R01GM127573). This work was partially supported by the National Nuclear Security Administration’s Office of Defense Nuclear Nonproliferation Research and Development. Work was performed in part at the National User Resource for Biological Accelerator Mass Spectrometry, which is operated at LLNL under the auspices of the U.S. Department of Energy under contract DE-AC52-07NA27344. The User Resource is supported by the National Institutes of Health, National Institute of General Medical Sciences under grant R24GM137748.

{\bf Author contributions:} Conceptualization: J.J. and A.D.M.; Experimental design: J.J. and A.D.M.; Data acquisition: J.J. and A.D.M.; Modeling: J.J.; Funding acquisition: A.D.M.;  Writing (original draft): J.J.; Writing (revision): J.J. and A.D.M.

{\bf Competing interests:} J.J. and A.D.M. are employees of Lawrence Livermore National Laboratory, which is managed by Lawrence Livermore National Security (LLNS) LLC. A patent application based on the 2C-CRDS technique, which is applied to the $^{14}$CO$_2$ measurements in this study, was filed by LLNS. The patent application was approved by the US Patent Office with Patent number US11585753.

{\bf Data and materials availability:} The data that support the findings of this study are available from the corresponding authors upon reasonable request.

\section*{Supplementary information}
Spectral fit\\
Details of the 2C-CRDS detection system\\
Sample preparation\\
Supplementary text\\
Figs. S1 to S5


\clearpage

\end{document}



\baselineskip24pt

\maketitle

\newpage
\section{Methods}

\subsection{Spectral fit}
\label{sec:fit}

Traditional lineshape functions such as the Voigt profile are inadequate for fitting the intra-cavity 2C peaks, which are in general asymmetric with respect to the pump detuning frequencies~\cite{mccartt2022room,jiang2021two}, e.g., the maximum of the $^{14}$CO$_2$ 2C transition from our current P(14)-P(17) pump-probe scheme occurs at a pump-detuning frequency of $\sim$30\,MHz. The $^{14}$CO$_2$ signals in Figs.~2a and 3a of the main text are modeled using the cavity-resonance-constrained Bloch equation formalism outlined in our previous papers~\cite{mccartt2022room,jiang2021two}. Considering the rapid rotational relaxation among various $J$-levels of $\nu_3=1$, a three-level system equation is effectively applied here to treat the four-level problem (scheme ii in Fig.~1c of the main text), i.e., the lower level of the probe transition, $\nu_3=1$ $J=17$, is assumed to be populated immediately after pump excitation of $\nu_3=1$ $J=13$. Limited by the current signal-to-noise ratios, the background 2C spectra from the $^{14}$C-depleted ``dead'' samples (i.e., ``Petrogenic gas'' and ``Coal'') are modeled as a sum of two Voigt functions. The homogenous and inhomogenous linewidths are assumed to have the same value for these two Voigt profile peaks.  The fit parameters for the background signals are determined from the ``Coal'' spectrum in Fig.~2a of the main text, and those for modeling the $^{14}$CO$_2$ 2C contribution are determined from a fit to the 2C-CRDS spectrum of a highly enriched $^{14}$CO$_2$ sample (40$\times$ natural abundance) after subtraction of the background 2C contribution. The observed 2C spectra from all other samples are modeled as a linear combination of the simulated $^{14}$CO$_2$ and dead spectra, by fitting only their respective amplitudes. The fit uncertainty for each data point ($\sigma$) is set according to the expected measurement precision, determined by the shot-to-shot 2C ringdown rate fluctuations ($\sigma_{sts}\sim5$\,s$^{-1}$) of the current detection system and the averaging time ($t_{avg}$), i.e., $\sigma=\sigma_{sts}/\sqrt{t_{avg}}$. The fitted model reproduces the experimental spectra with an average $\sigma$-weighted error close to unity (see Fig.~2d of the main text).

\subsection{Details of the 2C-CRDS detection system}
\label{sec:2CCRDS_details}
The three-mirror cavity consists of two plano mirrors and a plano-concave mirror with 1-m radius of curvature (LohnStar). The two plano mirrors are glued onto an invar cavity spacer. The concave mirror is housed in a piezoelectric-transducer (PZT) assembly, which is attached to the cavity spacer. The laser incidence angle at the PZT mirror is $\sim$1.5$^\circ$. The pump, probe, and reference lasers are continuous-wave QCL (Hamamatsu, HHL-package). The pump and probe lasers (1000\,mA maximum current) are each driven by a battery-powered QubeCL system from ppqSense. For both the pump and probe lasers (each modulated at 6\,MHz), light reflection off the cavity is measured with a HgCdTe (MCT) photodetector (Thorlabs PDAVJ8), and the MCT signal is demodulated with a frequency mixer (Mini-Circuit, ZRPD-1+). The resulting error signal is used as the input to a PID servo control loop (Vescent, D2-125-PL) to achieve laser frequency-locking to the cavity with the PDH method ($\sim$1\,MHz locking bandwidth). The probe cavity ringdown signals are measured by a liquid-nitrogen-cooled InSb detector (InfraRed Associates, Model IS-0.25) coupled to a pre-amplifier with 1\,MHz bandwidth (InfraRed Associates, INSB-1000). The intra-cavity pump power is stabilized based on the cavity-transmitted pump power, which is measured by an MCT detector (VIGO, PVI-4TE-6-1$\times$1/PIP-DC-20M-F-M8). Both the pump and probe signals are digitized on an oscilloscope (National Instrument, PXI-5922) that operates at 4\,Ms/s sampling rate and 20-bit analog input resolution.

The reference QCL is driven by a current controller from Wavelength Electronics (QCL500 Laboratory Series). The temperature of this QCL is regulated with a PI servo control loop (Wavelength Electronics, PTC2.5K-CH). After a double-pass through an optical cell (10 cm, 4.5 torr N$_2$O), the transmitted intensity of the reference laser (modulated at 3 MHz) is measured by an MCT photodetector (VIGO, PVI-4TE-6/PIP-DC-20M), and the signal is demodulated with a frequency mixer (Mini-Circuit, ZRPD-1+). The frequency of the reference laser is locked to the center of the $\nu_{3}=1-0$, R(16) transition of $^{15}$N$^{14}$N$^{16}$O at 2214.339 $\pm\:0.001$\,cm$^{-1}$ by a PI servo loop (New Focus, LB1005). The beatnote of the reference and pump lasers is measured by another MCT detector (VIGO PVI-4TE-10.6/FIP-1k-1G).  This beatnote provides frequency calibrations for the pump and probe lasers. In our experiments, the probe laser frequency is first roughly measured using a wavemeter (Bristol 771) and then assigned a frequency using the beatnote and the cavity mode spacing (using Eq.~1 in Ref \cite{jiang2021two}). Timing for the experiment is controlled with a custom code implemented on a field programable gate array (FPGA, National Instrument, PXIe-7976R and NI-5783). The FPGA system controls the AOMs (IntraAction Corp) for the pump and probe lasers, and provides corrections to the PDH servo.

\begin{figure}
\centering
\includegraphics[width=2.7 in]{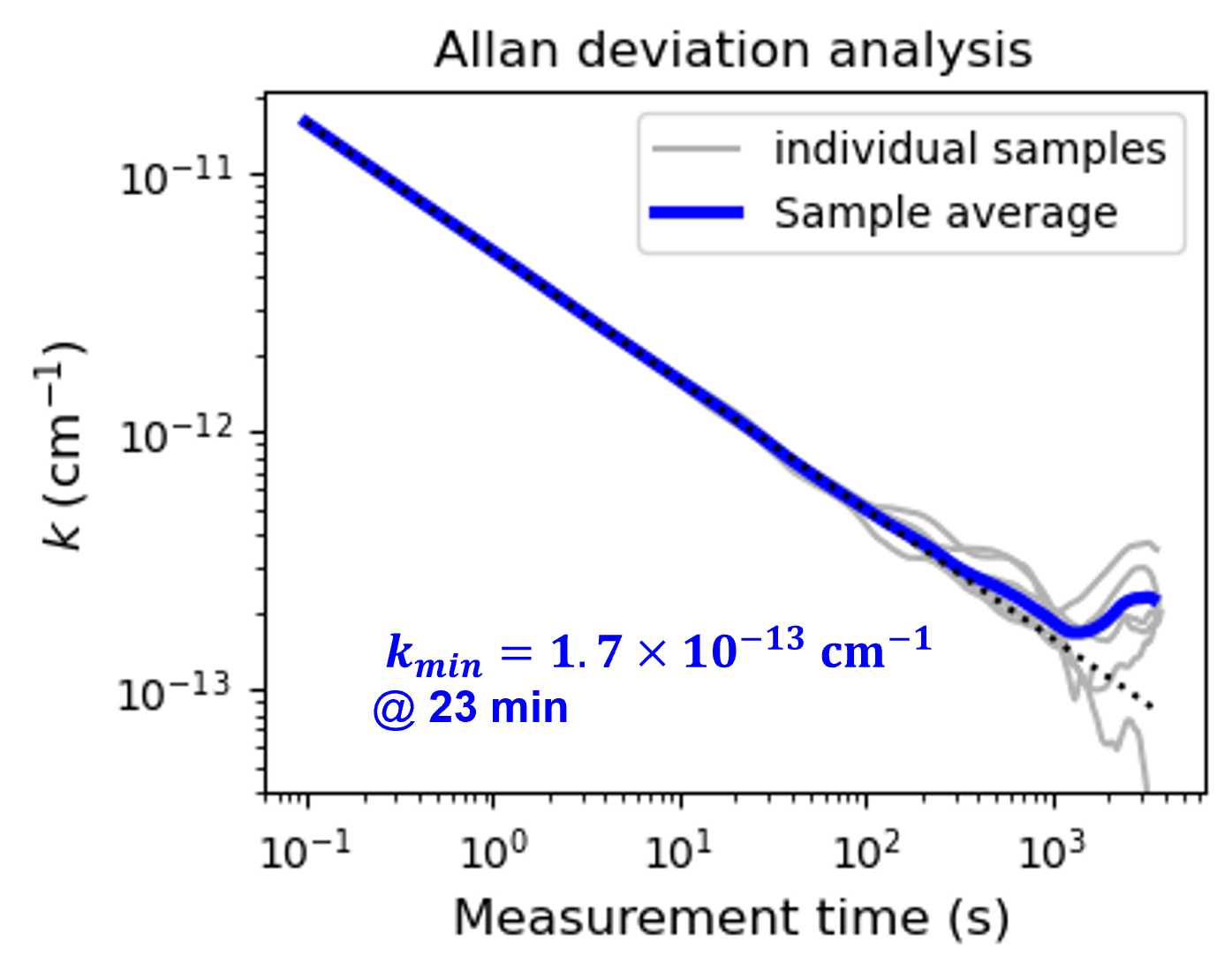}
\caption{Allan deviation analysis of 2C-CRDS measurements of multiple CO$_2$ samples (20.1\,torr) with $^{14}$C/C in the range of 0 to 1.81 ppt. The repetition rate of the probe ringdown events is $\sim$2.2\,kHz for all these measurements. The square root of the average of the Allan variances of individual sample measurements is used to determine the short-term sensitivity of the current 2C-CRDS setup. The dotted line indicates ideal averaging based on the standard error of the mean.}
\label{fig:allan}
\end{figure}

The Allan deviation analysis of 2C-CRDS signals from multiple $^{14}$CO$_2$ samples is provided in Fig.~\ref{fig:allan}. The minimum of the sample-averaged Allan deviation curve is used to determine the ``ultimate sensitivity'' of our current 2C-CRDS setup for $^{14}$CO$_2$ detection.

\section{Materials}

The sample preparation procedures of the $^{14}$C standard and bio-fuel samples are similar to those described in our previous work~\cite{mccartt2022room}. The ``Petrogenic gas'' sample is instrument-grade CO$_2$ from petroleum feedstock (Airgas). The other $^{14}$C samples used in this work are either in the solid ($^{14}$C standards) or liquid form (bio-fuels). The bio-fuel samples are transportation fuels containing 1-7$\%$ of biologically-derived carbon-containing materials. The ``Coal'' sample is composed of coal. The “Tiriwood” sample is “Belfast Pine, Sample B” from the Third International Radiocarbon Intercomparison with designator Q7780~\cite{scott2003part}. It was collected from a 5240-year-old tree (TIRI Wood, Pinus sylvestris). ``Ox1'' (Oxalic acid I, NIST designation SRM 4990 B) is a principle $^{14}$C standard, and was derived from a crop of 1955 sugar beet~\cite{stuiver1977discussion,mann1983international,stuiver1983international}. ``Ox2'' (Oxalic acid II, NIST designation SRM 4990 C) was derived from a crop of 1977 French beet molasses~\cite{stuiver1977discussion,mann1983international,stuiver1983international}. “IAEAC3” is cellulose produced in 1989 from $\sim$40 year old trees in Sweden~\cite{rozanski1992iaea}. ``ANU'' (IAEAC6) is sucrose produced from sugar cane grown between September 1965 and June 1971~\cite{rozanski1992iaea}. The ``ANU'' sample is named after the Australian National University, the home institute of H.A. Polach who prepared the first batch of the sample~\cite{polach20201}. The $^{14}$C contents in the $^{14}$C standard and bio-fuel samples have all been previously measured by AMS. 

Each of these samples is sealed in a quartz tube along with excess amount of copper oxide ($>$150\,mg), and is combusted at 900$^\circ$C for 2-4 hours. The quartz tube containing the CO$_2$ gas is cracked under vacuum inside a bellow tube attached to the gas manifold. The released gas first passes over an isopropanol/dry ice water trap. The gas sample is then exposed to a liquid nitrogen cold finger for removal of the non-condensible gas components (O$_2$ and N$_2$). The purified CO$_2$ gas is introduced to the optical cavity for 2C-CRDS measurements at 20.1 torr. 

\section{Supplementary text}
\label{sec:text}

\subsection{2C-CRDS signal variations}

\begin{figure}
\centering
\includegraphics[width=6.3 in]{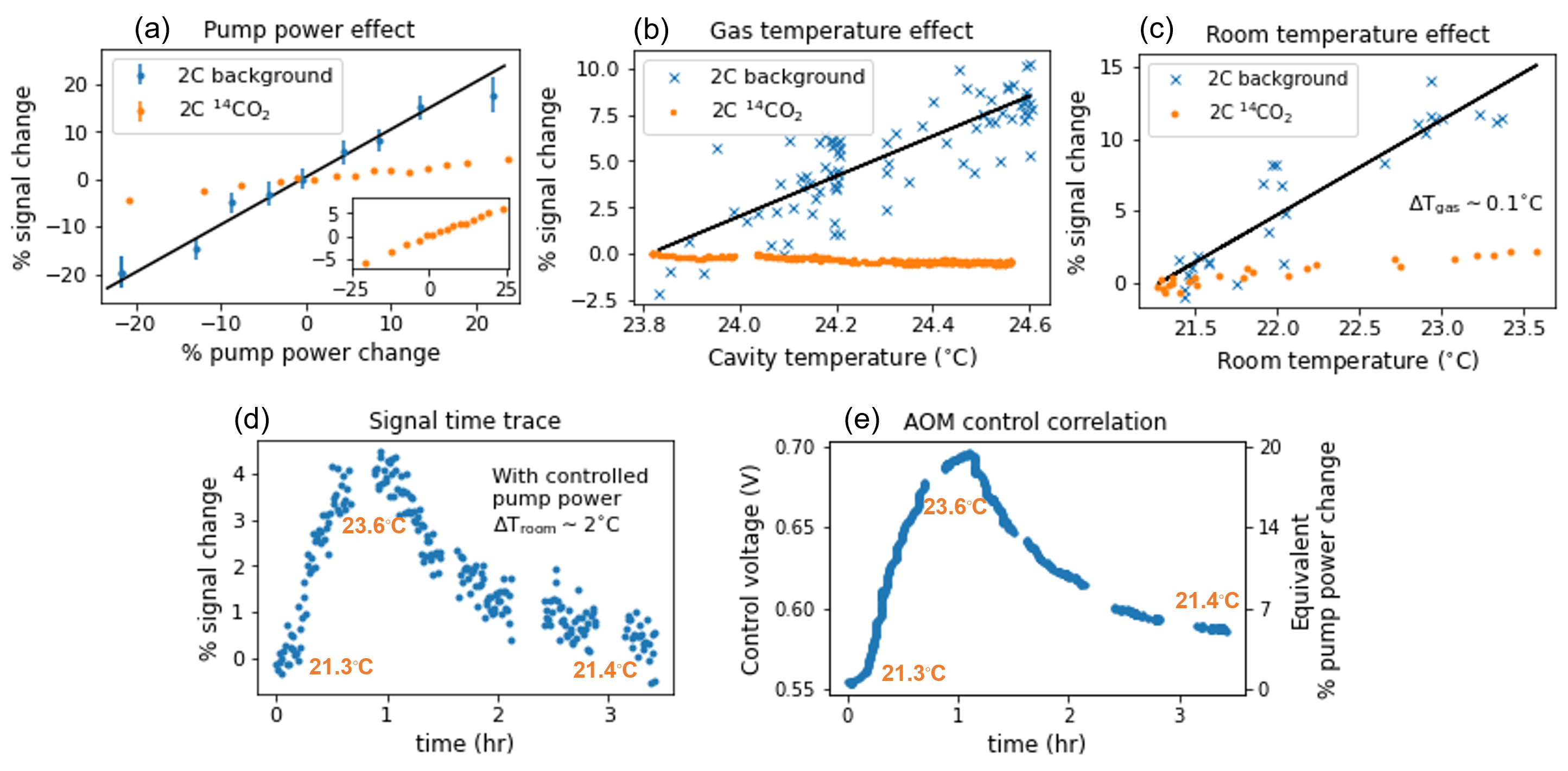}
\caption{Variations of 2C-CRDS signals due to changes of the experimental conditions. (a)-(c) Sensitivity of 2C signals from $^{14}$CO$_2$ and the background 2C process on various experimental conditions. (d) Time trace of 2C-CRDS signals from a $^{14}$CO$_2$ sample at 2$\times$ natural abundance concentration. (e) Changes of the AOM control voltage during the experiments in panel (d). This control voltage determines the pump power at the AOM output, and is automatically adjusted based on the photodiode measurement of the cavity-transmitted pump power. The $^{14}$CO$_2$ signals in panels (a), (c), and (d) are measured with the same $^{14}$CO$_2$ sample (2$\times$ natural abundance), and those in panel (b) are measured with a sample at 23$\times$ natural abundance concentration. The background 2C signals are all measured with a dead sample. Except for the inset in panel (a), an empirical linear dependence of the background 2C signals on pump power and temperature changes has been subtracted from the $^{14}$CO$_2$ signals in panels (a)-(c).}
\label{fig:environment}
\end{figure}

We have investigated the pump power and temperature dependence of the 2C-CRDS signals (Fig.~\ref{fig:environment}). In general, the 2C signals from $^{14}$CO$_2$ are significantly less sensitive to changes in the experimental conditions than the background 2C signals, which result from inadvertent hot-band excitation of other CO$_2$ isotopogoues. Among these observations, the strong dependence of the 2C signals on the ambient temperature is unexpected (Fig.~\ref{fig:environment}c). A 2$^\circ$C increase of the room temperature leads to $\sim$2$\%$ and $\sim$13$\%$ increase, respectively, for the 2C signals from $^{14}$CO$_2$ and the background 2C process. Note that this 2$^\circ$C change of the ambient temperature leads to only a modest increase in the temperature of the gas cavity ($T_{\textup{gas}}$$\sim$0.1$^\circ$C), which is housed inside a temperature-controlled experimental chamber. As can be inferred from Fig.~\ref{fig:environment}b, $>$1$^\circ$C increase of $T_{\textup{gas}}$ would be needed to cause the observed $\sim$13$\%$ increase of the background 2C signal in Fig.~\ref{fig:environment}c. Furthermore, the $^{14}$CO$_2$ 2C signal from a hotter gas sample slightly decreases (by $\sim$0.6$\%$ for a 0.8$^\circ$C increase of $T_{\textup{gas}}$) because of depletion of thermal population in the initial $\nu_3=0$, $J=14$ level of the pump-probe scheme. The observed ambient temperature dependence of the 2C signals in Fig.~\ref{fig:environment}c must be caused by temperature-induced changes in the experimental conditions external to the gas cavity. 

We believe that changes in the ambient temperature leads to mis-control of the intra-cavity pump power, which is stabilized based on the photodiode measurement of the cavity-transmitted pump power (see Fig.~1a of the main text and Section~S1B). As can be seen from Figs.~\ref{fig:environment}d and \ref{fig:environment}e, the temperature-related variations of the 2C-CRDS signals (from a $^{14}$CO$_2$ sample with 2$\times$ natural abundance concentration) correlate with changes of the control voltage for the pump laser power during the experiment. It appears that, as a result of the 2$^\circ$C increase of the room temperature, our current pump power controlling scheme inadvertently leads to nearly 20$\%$ increase for the pump laser power. Based on the pump power dependence of the 2C-CRDS signals from the same sample (see the inset of Fig.~\ref{fig:environment}a), this unexpected 20$\%$ variation of the pump power could be responsible for the observed $\sim$4$\%$ change of the 2C-CRDS signal in Fig.~\ref{fig:environment}d. 

We must point out that the applied changes to the experimental conditions in Fig.~\ref{fig:environment} are $\sim$10$\times$ larger than those during any of the $^{14}$C standards and bio-fuel measurements. As discussed in the main text, variations of the background 2C signals can be at least partially compensated by spectral fitting. It is more difficult to directly account for the much smaller but non-negligible variations of the $^{14}$CO$_2$ signals. For samples with close to natural $^{14}$C concentration, these variations could lead to $\sim$2\,ppq measurement error, given the stability of the experimental conditions in our laboratory. A more reliable and robust pump power stabilization scheme will be implemented in a future generation of the 2C-CRDS setup.

\subsection{Details of the intra-cavity pump-probe experiments on collision-induced population transfer}
The purpose of the pump-probe experiments on $^{13}$CO$_2$ in Fig.~4a of the main text is to gain insights into the collisional mechanisms responsible for the background 2C signals relevant to 2C-CRDS detection of $^{14}$CO$_2$. In each of these experiments, the probe laser wavenumber spans continuously from approximately 2213.88\,cm$^{-1}$ to 2214.13\,cm$^{-1}$. This continuous wavenumber coverage is achieved by splicing together 16 sectional spectra, each of which covers a frequency span of the cavity FSR (443.3\,MHz). During each sectional FSR scan, the pump radiation is coupled to the $same$ cavity resonance (i.e., with the same absolute mode index) that tunes across the resonance frequency of a target pump transition at approximately the mid-point of the FSR scan. This pump-probe frequency tuning scheme ensures that, during each probe sectional scan, the maximum absolute pump detuning frequency is less than $\frac{\textup{FSR}}{2}$ from the target $^{13}$CO$_2$ pump transition frequency. Both positive and negative 2C signals can occur depending on whether the pump-initiated population redistribution increases or decreases the population in the lower level of a specific probe transition.

\begin{figure}
\centering
\includegraphics[width=6.25 in]{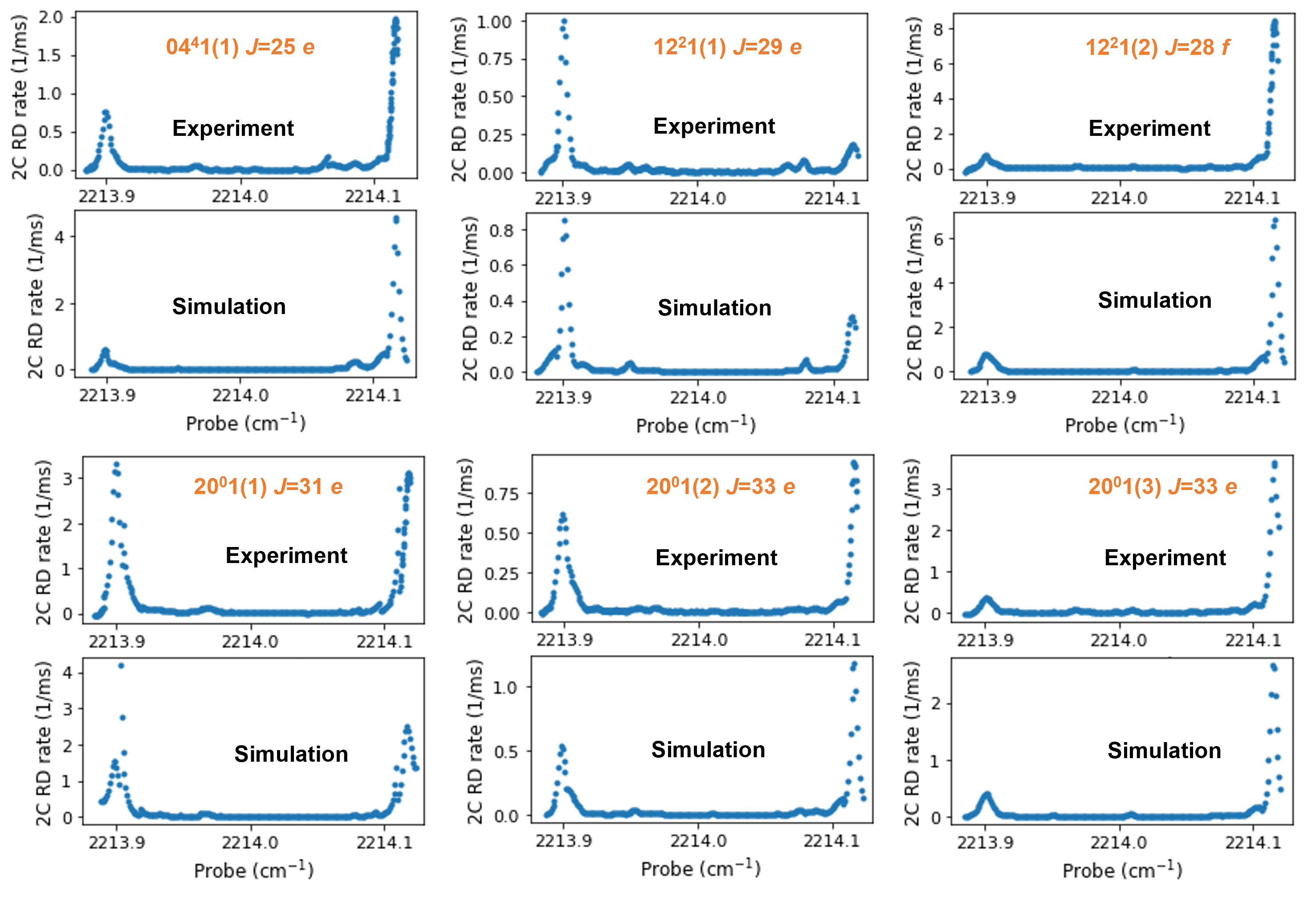}
\caption{Simulation of the observed collision-induced 2C spectra following pump excitation of six hot-band ro-vibrational transitions of $^{13}$CO$_2$. The six target $^{13}$CO$_2$ levels for the pump are indicated on the figures.}
\label{fig:col_sim}
\end{figure}

The intensities of the collision-induced 2C spectra are, in general, $not$ expected to be continuous across different sectional spectra, because of the discontinuity of the pump laser detuning frequency and thereby pump excitation probability at the splicing points of neighboring spectra. However, given that the resonance frequency of a target pump transition occurs, by design, close to the mid-point of each FSR scan in our experiments here, the pump excitation efficiency is largely equal at those splicing points. The intensities of the 2C signals due to excitation of a target pump transition are thus expected to be nearly continuous across neighboring spectra, which is largely found to be the case in Fig.~\ref{fig:col_sim}. Clear discontinuities are, however, present between some of the neighboring sectional spectra, in particular from the experiment for which $20^01(1)$ J=31$e$ is the target pump-populated $^{13}$CO$_2$ level (e.g., at the 2214.1 cm$^{-1}$ region of the corresponding spectra in Fig.~\ref{fig:col_sim}). The presence of these discontinuities indicates that at least part of the observed signals originate from a collision-induced four-level excitation process that does $not$ involve the target pump transition. The resonance frequency of the pump transition in that additional four-level excitation process could be far from the center frequency of the pump frequency tuning range designed for the original target pump transition. For example, 2C signals due to far-off-resonant pump excitation of the $11^11(2)\leftarrow11^10(2)$ P(42$f$) transition of $^{13}$CO$_2$ ($>$500\,MHz detuned) are believed to be the cause of the intensity jumps between neighboring spectra at the 2214.1 cm$^{-1}$ region of the ``$20^01(1)$ J=31$e$'' experiment in Fig.~\ref{fig:col_sim}. As can be seen from Fig.~\ref{fig:col_sim}, the non-continuous 2C spectra from that experiment are well reproduced by our simulation, which takes into account all possible pump excitation transitions within 8\,GHz of the resonance frequency of a chosen target pump transition. 


\subsection{Modeling the collision-induced background signals}
\label{app:collisions}

\begin{figure}
\centering
\includegraphics[width=2.15 in]{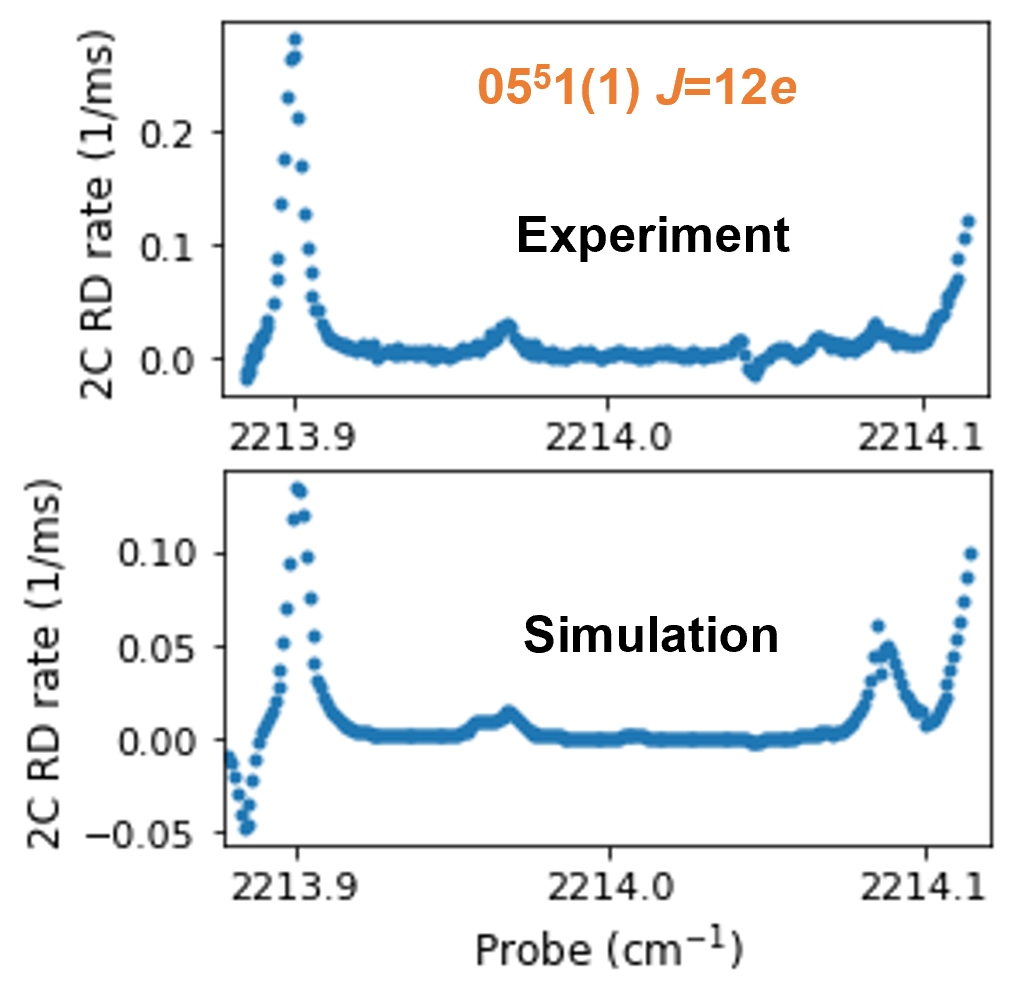}
\caption{Simulation of the observed collision-induced 2C spectra following pump excitation of the $05^51(1)\leftarrow05^50(1)$ P(13) transition of $^{13}$CO$_2$.}
\label{fig:col_sim_05511}
\end{figure}

\begin{figure}
\centering
\includegraphics[width=6.3 in]{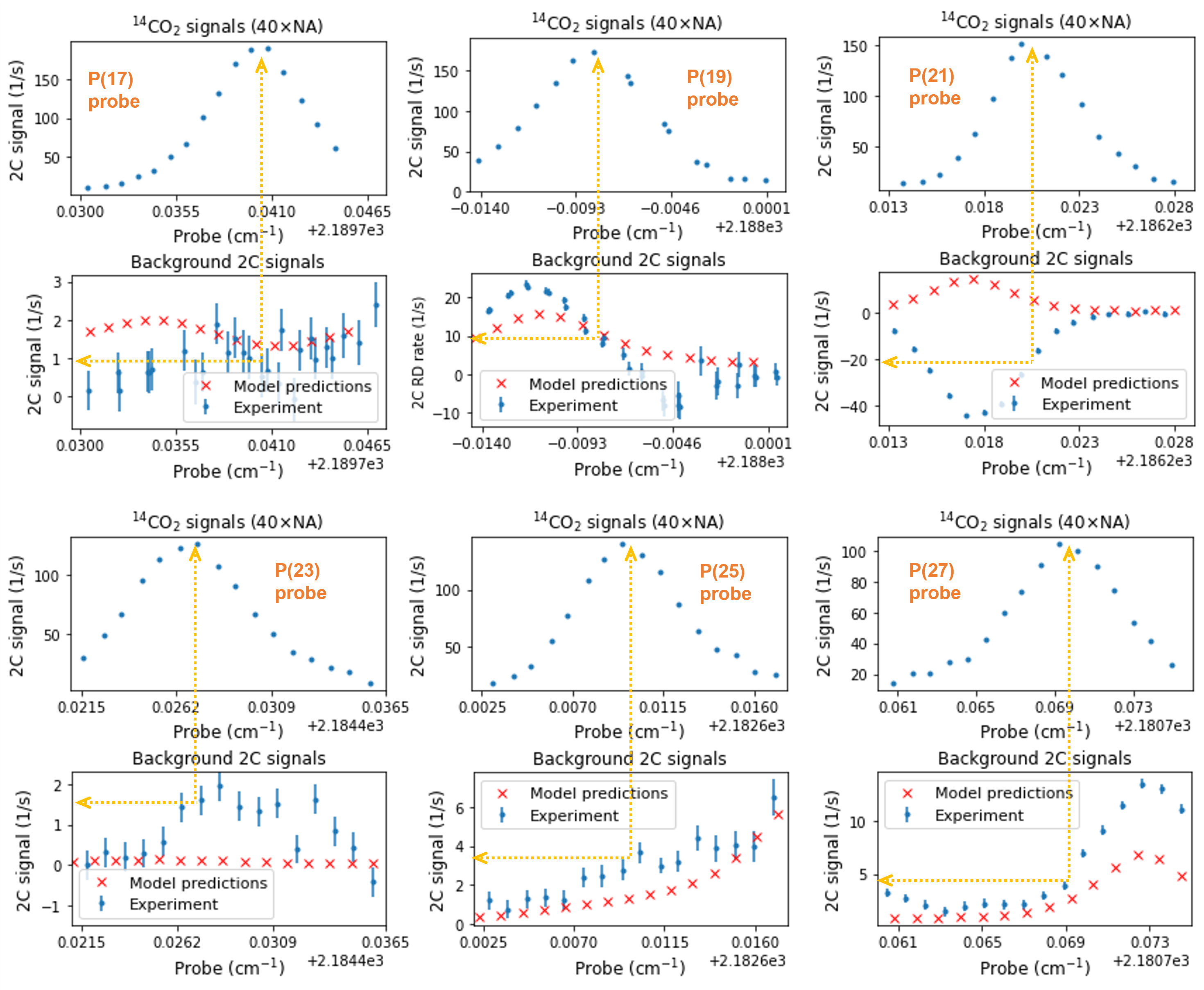}
\caption{Investigation of collision-induced background 2C signals at the resonance frequency regions of various $\nu_3=2\leftarrow1$, P-branch probe transitions of $^{14}$CO$_2$ (with the same $\nu_3=1\leftarrow0$, P(14) pump). The predicted background 2C spectra are based on the collision model derived from the pump-probe experiments shown in Fig.~4a of the main text. The background 2C signals are measured with a ``Petrogenic gas'' (dead) sample. The observed P-branch probe spectra from a $^{14}$CO$_2$ sample at 40$\times$ natural abundance concentration are also shown here. The double-headed right-angle arrows are used to indicate the background 2C signal levels at the resonance frequency of these P-branch transitions.}
\label{fig:col_sim_P}
\end{figure}

The collision model for the background 2C signals described in the main text is applied to simulate the collision-induced 2C spectra from all possible isotopologues of CO$_2$, based on the linelist in Ref \cite{huang2022ames}. The simulated spectra for different isotopologues, weighted by their natural abundances used in Ref \cite{gordon2022hitran2020}, are combined to yield the simulated spectra. The model suggests that the observed 2C signals in Fig.~4a of the main text and Figs.~\ref{fig:col_sim}-\ref{fig:col_sim_P} here originate predominantly from the $^{13}$C$^{16}$O$_2$ species, although non-negligible contributions from $^{13}$C$^{16}$O$^{18}$O are present in some of the spectra. 

In the model, the ``$J$-changing collisions'' pathway (Pathway 1) is treated as a special case of Pathway 2, for which the vibrational energy difference is zero for Pathway 1. The intensity of the collision-induced signal is assumed to be dependent on the vibrational energy gap for the responsible collisional pathway. For simplicity, the simulated 2C signals decrease by 10$\times$ for every 200 cm$^{-1}$ increase in $|\Delta E|$. Furthermore, the effects on the collisional transfer rates due to differences in the basis state composition among different Fermi polyads and polyad members are ignored in our model. The collision-induced homogenous linewidths are taken to be 4\,MHz/torr (half-width at half maximum) for all the probe transitions. Their inhomogenous linewidths are fixed to the room-temperature value. 

Under these assumptions, we qualitatively reproduce the observed collision-induced 2C spectra in Fig.~\ref{fig:col_sim} by adjusting an overall intensity scaling factor and the power-broadened linewidth for the pump transitions (assumed to be the same for all pump transitions used in this work). Initially, we assume that the rotational distribution is thermal (300\,K) for all the pump-populated vibrational states in the model. Later, we find that by further assuming a weak $\Delta E$-dependence for rotational relaxation, in the form of the exponential model of Polanyi and Woodall~\cite{polanyi1972mechanism,yardley2012introduction} (i.e., $\propto e^{-c \Delta E}$, where $c=0.00355$\,(cm$^{-1}$)$^{-1}$), the simulation achieves better agreement with the experimental observations.

The target pump-populated vibrational state in Fig.~\ref{fig:col_sim_05511}, 05$^5$1(1), has one more quantum of excitation in $\nu_2$ than those in Fig.~\ref{fig:col_sim}. Following pump excitation into the six target $^{13}$CO$_2$ levels in Fig.~\ref{fig:col_sim}, only one quantum exchange of the $\nu_2$ mode vibration with the $^{12}$CO$_2$ bath is required to observe the probe transition at 2214.12 cm$^{-1}$. However, an overall two-quantum exchange is needed to observe the same probe signal from 05$^5$1(1). The probe spectrum from this higher-energy 05$^5$1(1) state in Fig.~\ref{fig:col_sim_05511} thus allows us to estimate the effect of exchanging additional quanta of $\nu_2$ on the observed probe signal intensities. The simulated spectra in Figs.~\ref{fig:col_sim_05511} and \ref{fig:col_sim_P} are generated by assuming a 10$\times$ reduction in the probe signal for each additional $\nu_2$ exchange. 

Note that, our CO$_2$-transition-based model fails to predict the $negative$ 2C peak from the P(21) probe spectrum in Fig.~\ref{fig:col_sim_P}. This observed negative signal can be explained by a ``$V$-type'' four-level excitation of the $^{14}$N$_2$$^{18}$O contaminant in the gas sample, i.e., $\nu_3=1\leftarrow0$, P(3) pump (with -1\,GHz detuning) and  $\nu_3=1\leftarrow0$, P(34) probe. 

\clearpage

\printbibliography